\documentclass[12pt,subeqn]{iopart}

\usepackage{iopams}
\usepackage{setstack}
\usepackage{graphicx}
\usepackage{epstopdf}
\begin{document}

\title[Regularized linearization for quantum nonlinear optical cavities]
{Regularized linearization for quantum nonlinear optical cavities:
Application to Degenerate Optical Parametric Oscillators}

\author{Carlos Navarrete--Benlloch$^1$, Eugenio Rold\'an$^{1,2}$, Yue Chang$^1$ and Tao Shi$^1$}

\address{$^1$ Max-Planck-Institut f\"{u}r Quantenoptik, Hans-Kopfermann-strasse 1, 85748 Garching, Germany}
\address{$^2$ Departament d'\`Optica, Universitat de Val\`encia, Dr. Moliner 50, 46100 Burjassot, Spain}
\eads{\mailto{carlos.navarrete@mpq.mpg.de}, \mailto{eugenio.roldan@uv.es}, \mailto{yue.chang@mpq.mpg.de}, \mailto{tao.shi@mpq.mpg.de}}

\begin{abstract}
Nonlinear optical cavities are crucial both in classical and quantum optics;
in particular, nowadays optical parametric oscillators are one of the most
versatile and tunable sources of coherent light, as well as the sources of the
highest quality quantum-correlated light in the continuous variable regime.
Being nonlinear systems, they can be driven through critical points in which a
solution ceases to exist in favour of a new one, and it is close to these
points where quantum correlations are the strongest. The simplest description
of such systems consists in writing the quantum fields as the classical part
plus some quantum fluctuations, linearizing then the dynamical equations with
respect to the latter; however, such an approach breaks down close to critical
points, where it provides unphysical predictions such as infinite photon
numbers. On the other hand, techniques going beyond the simple linear
description become too complicated especially regarding the evaluation of
two-time correlators, which are of major importance to compute observables
outside the cavity. In this article we provide a regularized linear
description of nonlinear cavities, that is, a linearization procedure yielding
physical results, taking the degenerate optical parametric oscillator as the
guiding example. The method, which we call self-consistent linearization, is
shown to be equivalent to a general Gaussian ansatz for the state of the
system, and we compare its predictions with those obtained with available
exact (or quasi-exact) methods.
\end{abstract}

\pacs{42.65.Sf, 42.65.Yj, 42.50.Lc, 42.50.Ex}

\maketitle

\section{Introduction}

Nonlinear optical cavities, that is, cavities containing some element whose
response to an applied optical field is nonlinear, are very important both in
classical and quantum optics. In the classical domain they allow for effects
such as frequency conversion \cite{Boyd} or spatio-temporal pattern formation
\cite{PatternFormation}, while in the quantum domain they allow for the
generation of quantum correlations manifesting as squeezing or entanglement
\cite{BlueBook}, basic resources for modern applications such as
high-precission measurements \cite{Goda08, Vahlbruch05, Treps03, Treps02} and
quantum information communication and processing \cite{Braunstein05,
Weedbrook12}. The paradigmatic example of such systems are degenerate optical
parametric oscillators (DOPOs); these consist in a resonator containing a
crystal with second-order nonlinearity, which, when pumped with an external
monochromatic laser, is able to generate photons at the subharmonic frequency
through the process known as parametric down-conversion (PDC). The interplay
between the nonlinear parametric amplification and the cavity losses sets a
threshold power below which no subharmonic field is generated at the classical
level, and it is close to this critical point where quantum effects are the
largest; in particular, more than 90\% of quadrature squeezing has been
experimentally generated with such a system \cite{Eberle10, Mehmet10,
Vahlbruch08, Takeno07}.

In order to analyze the quantum properties of these systems, the simplest and
most widely used technique consists in expanding the field as a classical part
plus some quantum fluctuations, and linearize the dynamical equations with
respect to the latter \cite{Drummond80, Lugiato81}; from another (equivalent)
point of view, this approximation means that the state is taken to be
Gaussian, with a mean coinciding with the classical field amplitude, and a
covariance matrix accounting for the quantum fluctuations. The problem with
such approximation is that it gives unphysical predictions such as perfect
quadrature squeezing (which requires infinite energy) close to the critical
points of the classical theory \cite{Collett84}.

Especially for DOPOs, people have developed more refined techniques which go
beyond the linear approximation, correcting this unphysical predictions for
quantum correlations. Among these techniques, the ones based on the positive
$P$ representation \cite{PositiveP} are of especial relevance; this
representation allows for an exact mapping of the quantum dynamics onto a
set of classical stochastic equations from which a proper perturbation
expansion or numerical simulation can be carried even at the critical point
\cite{Kinsler93, Kinsler95, Drummond02, Chaturvedi99, Pope00}. Moreover, in
the limit in which the pump dynamics can be adiabatically eliminated, exact
solutions to the steady-state positive $P$ distribution of the DOPO are known
\cite{Drummond80, Wolinsky88, WallsMilburnBook}. The DOPO dynamics close to
the critical point has even been analyzed via non-equilibrium many-body
techniques such as the Keldysh formalism \cite{MertensPRL93, MertensPRA93,
Veits97}.

The problem with all these beyond-linear techniques is that they are quite
complicated when it comes to the evaluation of two-time correlation functions
needed for predictions concerning measurements outside the cavity, functions
which, on the other hand, are straightforwardly evaluated within a linearized
or Gaussian description. Motivated especially by this last fact, in this work
we offer a linear theory (or, equivalently, a Gaussian ansatz for the state of
the system) which regularizes the unphysical predictions offered by the usual
linearization procedure.

The article is organized as follows. In the next section we describe the
quantum model for DOPOs in the Schr\"{o}dinger and Heisenberg pictures, using,
respectively, the cavity modes' master equation and a set of quantum Langevin
equations. Then, in Section \ref{Heisenberg} we introduce our regularized self-consistent linearization procedure in the Heisenberg picture, to show in
Section \ref{Schrodinger} that it is completely equivalent to using a general
Gaussian ansatz for the state in the Schr\"{o}dinger picture. In Section
\ref{QuantumResults} we compare the quantum correlations (quadrature
fluctuations) obtained through this method with previous exact (or
quasi-exact) methods, showing that, despite its simplicity, it agrees with the
latter not only qualitatively, but also quite well quantitatively. In the last
section we conclude and comment on other systems where the method could be applied.

\section{The DOPO model}

We consider a cavity containing a $\chi^{(2)}$-crystal, pumped with a laser
resonant with a cavity mode at frequency $2\omega_{0}$ (\textit{pump} mode),
such that photons can be down-converted inside the crystal to the subharmonic
resonance $\omega_{0}$ (\textit{signal} mode). Denoting by $\hat
{a}_{\mathrm{p}}$ and $\hat{a}_{\mathrm{s}}$ the annihilation operators for
pump and signal photons, respectively, the (interaction picture) Hamiltonian
which describes such scenario is given by $\hat{H}_{\mathrm{DOPO}}=\hat
{H}_{\mathrm{inj}}+\hat{H}_{\mathrm{PDC}}$ \cite{Drummond80,
Lugiato81,Collett84}, with%
\begin{equation}
\hat{H}_{\mathrm{inj}}  =\mathrm{i}\hbar\mathcal{E}_{\mathrm{p}}(\hat
{a}_{\mathrm{p}}^{\dagger}-\hat{a}_{\mathrm{p}}), \qquad \hat{H}_{\mathrm{PDC}}  =\mathrm{i}\hbar\frac{\chi}{2}(\hat{a}_{\mathrm{p}%
}\hat{a}_{\mathrm{s}}^{\dagger2}-\hat{a}_{\mathrm{p}}^{\dagger}\hat{a}_{\mathrm{s}}^{2}),
\end{equation}
where $\mathcal{E}_{\mathrm{p}}$ is proportional to the amplitude of the
injected laser (whose phase is taken as a reference for any other, what allows
taking this parameter as real), and $\chi$ is proportional to the nonlinear
susceptibility of the crystal as well as the overlapping between the spatial
modes involved in the down-conversion process. In addition to these coherent
processes, we need to introduce the cavity losses; there are two different
ways in which this can be done. In the Schr\"{o}dinger (interaction-)picture,
in which the state of the system $\hat{\rho}$ evolves while operators are
fixed, such irreversible processes are accounted for by extra terms in the
\textit{master equation} as \cite{GardinerZollerBook, WallsMilburnBook}%
\begin{equation}
\frac{d\hat{\rho}}{dt}=\left[  \frac{\hat{H}_{\mathrm{DOPO}}}{\mathrm{i}\hbar
},\hat{\rho}\right]  +\sum_{j=\mathrm{p},\mathrm{s}}\gamma_{j}(2\hat{a}%
_{j}\hat{\rho}\hat{a}_{j}^{\dagger}-\hat{a}_{j}^{\dagger}\hat{a}_{j}\hat{\rho
}-\hat{\rho}\hat{a}_{j}^{\dagger}\hat{a}_{j}), \label{MasterEq}%
\end{equation}
where the damping rates $\gamma_{j}$ are proportional to the transmissivity of
the coupling mirror at the corresponding frequency.

On the other hand, in the Heisenberg (interaction-)picture, the bosonic operators evolve according to
the \textit{quantum Langevin equations }\cite{GardinerZollerBook,
WallsMilburnBook}
%
\begin{eqnarray}
\frac{d\hat{a}_{\mathrm{p}}}{dt}  &  =\mathcal{E}_{\mathrm{p}}-\gamma
_{\mathrm{p}}\hat{a}_{\mathrm{p}}-\frac{\chi}{2}\hat{a}_{\mathrm{s}}^{2}%
+\sqrt{2\gamma_{\mathrm{p}}}\hat{a}_{\mathrm{p,in}}(t), \label{Qlang}
\\
\frac{d\hat{a}_{\mathrm{s}}}{dt}  &  =-\gamma_{\mathrm{s}}\hat{a}_{\mathrm{s}%
}+\chi\hat{a}_{\mathrm{p}}\hat{a}_{\mathrm{s}}^{\dagger}+\sqrt{2\gamma
_{\mathrm{s}}}\hat{a}_{\mathrm{s,in}}(t), \nonumber
\end{eqnarray}
%
in which the \textit{input} operators satisfy correlations%
\begin{equation}
\left\langle \hat{a}_{j\mathrm{,in}}(t)\right\rangle =\left\langle \hat
{a}_{j\mathrm{,in}}(t)\hat{a}_{l\mathrm{,in}}(t^{\prime})\right\rangle
=0,\quad\left\langle \hat{a}_{j\mathrm{,in}}(t)\hat{a}_{l\mathrm{,in}%
}^{\dagger}(t^{\prime})\right\rangle =\delta_{jl}\delta(t-t^{\prime}),
\label{InCorr}%
\end{equation}
and account for the vacuum driving fields entering the cavity through the
partially transmitting mirror.

In the following we explain our regularized linearization of this nonlinear
system in both pictures, since they provide different intuitive ideas of what
the procedure means.

\section{Heisenberg picture approach: self-consistent
linearization\label{Heisenberg}}

Before explaining the procedure, let us define the following dimensionless
parameters%
\begin{equation}
\sigma=\mathcal{E}_{\mathrm{p}}\chi/\gamma_{\mathrm{p}}\gamma_{\mathrm{s}%
},\quad\kappa=\gamma_{\mathrm{p}}/\gamma_{\mathrm{s}},\quad%
g=\chi/\sqrt{\gamma_{\mathrm{p}}\gamma_{\mathrm{s}}}, \label{NormPars}%
\end{equation}
and normalized variables%
\begin{equation}
\tau=\gamma_{\mathrm{s}}t,\quad\hat{b}_{\mathrm{s}}=g\hat{a}_{\mathrm{s}%
},\quad\hat{b}_{\mathrm{p}}=\sqrt{\kappa}g\hat{a}_{\mathrm{p}},\quad\hat{b}_{j\mathrm{,in}}(\tau)=\gamma_{\mathrm{s}}^{-1/2}\hat{a}%
_{j\mathrm{,in}}(\gamma_{\mathrm{s}}^{-1}\tau), \label{NormVars}%
\end{equation}
in terms of which the quantum Langevin equations \eref{Qlang} are rewritten
as
%
\begin{eqnarray}
\frac{1}{\kappa}\frac{d\hat{b}_{\mathrm{p}}}{d\tau}  &  =\sigma-\hat
{b}_{\mathrm{p}}-\frac{1}{2}\hat{b}_{\mathrm{s}}^{2}+\sqrt{2}g\hat
{b}_{\mathrm{p,in}}(\tau), \label{NormLangevin}
\\
\frac{d\hat{b}_{\mathrm{s}}}{d\tau}  &  =-\hat{b}_{\mathrm{s}}+\hat
{b}_{\mathrm{p}}\hat{b}_{\mathrm{s}}^{\dagger}+\sqrt{2}g\hat{b}_{\mathrm{s,in}%
}(\tau); \nonumber
\end{eqnarray}
%
note that the normalized input operators satisfy the correlations
(\ref{InCorr}) but now with respect to the new dimensionless time $\tau$.

\begin{figure}[t]
\flushright \includegraphics[width=\textwidth]{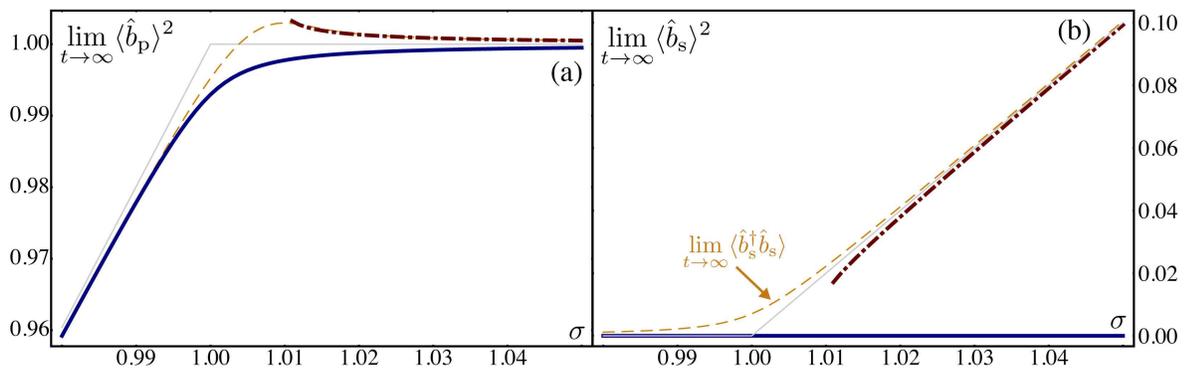}
\caption{Mean-field steady-state intensities of the pump (a) and signal (b)
modes as a function of the injection parameter $\sigma$. We have chosen
$\kappa=1$ and $g=0.01$, but similar figures are found for any other choice of
these parameters. The thin-solid light-grey curve corresponds to the usual
linearized description, while the solid blue line and dashed-dotted red line
correspond to our self-consistent method below and above threshold,
respectively. In (a) the dashed yellow line corresponds to the predictions of
Drummond and collaborators' perturbative analysis \cite{Kinsler95, Drummond02}
(see Section \ref{QuantumResults}), while in (b) it corresponds to the number
of (normalized) signal photons obtained from our self-consistent
below-threshold solution, showing how it is not divergent at threshold, in
contrast with the predictions found with the usual linearization approach.}
\label{Fig1ab}
\end{figure}

In order to linearize these equations, we expand the annihilation operators as
$\hat{b}_{j}=\beta_{j}+\delta\hat{b}_{j}$, with $\beta_{j}$ some amplitudes
which one identifies with the \textit{mean-field} part of the modes
$\langle\hat{b}_{j}\rangle$, and $\delta\hat{b}_{j}$ the operators accounting
for \textit{quantum fluctuations} around such mean-field, with respect to
which the theory will be linearized. The mean-field amplitudes can be
evaluated by taking the quantum expectation value of the quantum Langevin
equations \eref{NormLangevin}, leading to
%
\begin{eqnarray}
\frac{1}{\kappa}\frac{d\beta_{\mathrm{p}}}{d\tau}  &  =\sigma-\beta
_{\mathrm{p}}-\frac{1}{2}\beta_{\mathrm{s}}^{2}-\frac{1}{2}\langle\delta
\hat{b}_{\mathrm{s}}^{2}\rangle,\label{MFeqs}
\\
\frac{d\beta_{\mathrm{s}}}{d\tau}  &  =-\beta_{\mathrm{s}}+\beta_{\mathrm{p}%
}\beta_{\mathrm{s}}^{\ast}+\langle\delta\hat{b}_{\mathrm{p}}\delta\hat
{b}_{\mathrm{s}}^{\dagger}\rangle. \nonumber
\end{eqnarray}
%
In the usual approach, these mean-field equations are solved by assuming that
the state is coherent in all modes, hence neglecting the $\langle\delta\hat
{b}_{\mathrm{s}}^{2}\rangle$ and $\langle\delta\hat{b}_{\mathrm{p}}\delta
\hat{b}_{\mathrm{s}}^{\dagger}\rangle$\ terms, what gives rise to the
classical equations that would have been obtained from Maxwell's equations; in
other words, the mean-field amplitudes $\beta_{j}$ are taken to be the
classical solutions of the system \cite{Drummond80, Lugiato81}. In particular,
in the case of equations \eref{MFeqs}, this coherent mean-field ansatz
provides two different types of steady-state solutions depending on the
injection parameter $\sigma$ (see the thin-solid light-grey curve of Figure \ref{Fig1ab}): one known as the \textit{below-threshold
solution} with $(\bar{\beta}_{\mathrm{s}}=0,\bar{\beta}_{\mathrm{p}}=\sigma)$
which is the only stable solution for $\sigma\leq1$, and another (bistable)
solution for $\sigma>1$ known as the \textit{above-threshold solution} with
the signal field switched on $(\bar{\beta}_{\mathrm{s}}=\pm\sqrt{\sigma
-1},\bar{\beta}_{\mathrm{p}}=1)$. The injection $\sigma=1$ marks a
\textit{critical point} where the below-threshold solution changes from stable
to unstable, and, as we argue below, this sudden change in the stability
conditions is what generates unphysical predictions in the linear theory
\cite{Drummond80, Lugiato81,Collett84}.  

In order to correct such a problem, we propose to incorporate some information
of the quantum dynamics in the quantum-fluctuations-dependent mean-field
equations (\ref{MFeqs}), with the purpose of obtaining a better ansatz for the
mean-field amplitudes. Concretely, using (\ref{MFeqs}), the quantum Langevin
equations are rewritten in terms of the fluctuations $\delta\hat{b}_{j}$ as%
%
\begin{eqnarray}
\frac{1}{\kappa}\frac{d\delta\hat{b}_{\mathrm{p}}}{d\tau}  &  =-\delta\hat
{b}_{\mathrm{p}}-\frac{1}{2}\beta_{\mathrm{s}}\delta\hat{b}_{\mathrm{s}}%
-\frac{1}{2}\left(  \delta\hat{b}_{\mathrm{s}}^{2}-\langle\delta\hat
{b}_{\mathrm{s}}^{2}\rangle\right)  +\sqrt{2}g\hat{b}_{\mathrm{p,in}}(\tau),
\\
\frac{d\delta\hat{b}_{\mathrm{s}}}{d\tau}  &  =-\delta\hat{b}_{\mathrm{s}%
}+\beta_{\mathrm{p}}\delta\hat{b}_{\mathrm{s}}^{\dagger}+\beta_{\mathrm{s}%
}^{\ast}\delta\hat{b}_{\mathrm{p}}+\left(  \delta\hat{b}_{\mathrm{p}}%
\delta\hat{b}_{\mathrm{s}}^{\dagger}-\langle\delta\hat{b}_{\mathrm{p}}%
\delta\hat{b}_{\mathrm{s}}^{\dagger}\rangle\right)  +\sqrt{2}g\hat
{b}_{\mathrm{s,in}}(\tau). \nonumber
\end{eqnarray}
%
We can proceed then as in the regular linearization by neglecting the
nonlinear fluctuations $\delta\hat{b}_{\mathrm{s}}^{2}-\langle\delta\hat
{b}_{\mathrm{s}}^{2}\rangle$ and $\delta\hat{b}_{\mathrm{p}}\delta\hat
{b}_{\mathrm{s}}^{\dagger}-\langle\delta\hat{b}_{\mathrm{p}}\delta\hat
{b}_{\mathrm{s}}^{\dagger}\rangle$, obtaining the linear system%
\begin{equation}
\frac{d}{d\tau}\delta\mathbf{\hat{b}}=\mathcal{L}(\beta_{\mathrm{s}}%
,\beta_{\mathrm{p}})\delta\mathbf{\hat{b}}+\sqrt{2}g\mathbf{\hat{b}%
}_{\mathrm{in}}(\tau), \label{LinLan}%
\end{equation}
with $\delta\mathbf{\hat{b}}=\mathrm{col}(\delta\hat{b}_{\mathrm{p}},\delta\hat
{b}_{\mathrm{p}}^{\dag},\delta\hat{b}_{\mathrm{s}},\delta\hat{b}_{\mathrm{s}}^{\dag})$,
$\mathbf{\hat{b}}_{\mathrm{in}}=\mathrm{col}(\kappa\hat{b}%
_{\mathrm{p,in}},\kappa\hat{b}_{\mathrm{p,in}}^{\dag},\hat{b}_{\mathrm{s,in}%
},\hat{b}_{\mathrm{s,in}}^{\dag})$, and where the so-called \textit{linear
stability matrix} is defined as%
\begin{equation}
\mathcal{L=}\left(
\begin{array}
[c]{cccc}%
-\kappa & 0 & -\kappa\beta_{\mathrm{s}} & 0\\
0 & -\kappa & 0 & -\kappa\beta_{\mathrm{s}}^{\ast}\\
\beta_{\mathrm{s}}^{\ast} & 0 & -1 & \beta_{\mathrm{p}}\\
0 & \beta_{\mathrm{s}} & \beta_{\mathrm{p}}^{\ast} & -1
\end{array}
\right)  ; \label{L}%
\end{equation}
the difference now is that, instead of substituting the classical solutions
which give rise to a singular stability matrix at threshold (what can be
traced as the source of all the unphysical predictions\footnote{In particular,
the singular character of the linear stability matrix occurs because one of
its eigenvalues becomes zero at $\sigma=1$ (it changes from negative
to positive, corresponding to the destabilization of the below-threshold
solution when crossing the threshold); on the other hand, it is clear that
within this linear description, the correlators of quantum fluctuations are
inversely proportional to combinations of these eigenvalues, and hence some
might diverge at threshold, what we will show later explicitly.}), the
mean-field amplitudes $\beta_{j}$ are left as unknown variables which are
found self-consistently by calculating the $\langle\delta\hat{b}_{\mathrm{s}%
}^{2}\rangle$ and $\langle\delta\hat{b}_{\mathrm{p}}\delta\hat{b}_{\mathrm{s}%
}^{\dagger}\rangle$\ terms from this linear system, and plugging them into the
mean-field equations (\ref{MFeqs}). In this case, it is simple but lengthy,
for example by finding the (bi-orthonormal) eigensystem of (\ref{L}), to
obtain the following steady-state expressions for these correlators
%
\begin{eqnarray}
\lim_{\tau\rightarrow\infty}\langle\delta\hat{b}_{\mathrm{s}}^{2}\rangle &
=-\frac{g^{2}\beta_{\mathrm{p}}\left[  I_{\mathrm{p}}-(1+\kappa
)(1+I_{\mathrm{s}})(1+\kappa+I_{\mathrm{s}})\right]  }{2\left[  I_{\mathrm{p}%
}-(1+\kappa)^{2}\right]  \left[  I_{\mathrm{p}}-(1+I_{\mathrm{s}})^{2}\right]
},\label{SS-moments}
\\
\lim_{\tau\rightarrow\infty}\langle\delta\hat{b}_{\mathrm{p}}\delta\hat
{b}_{\mathrm{s}}^{\dagger}\rangle &  =-\frac{g^{2}\kappa\beta_{\mathrm{s}%
}I_{\mathrm{p}}(2+\kappa+I_{\mathrm{s}})}{2\left[  I_{\mathrm{p}}%
-(1+\kappa)^{2}\right]  \left[  I_{\mathrm{p}}-(1+I_{\mathrm{s}})^{2}\right]
}, \nonumber
\end{eqnarray}
%
where we have introduced what we call the mean-field \textit{intensities}
$I_{j}=\left\vert \beta_{j}\right\vert ^{2}$, complemented with the
\textit{phases} $\varphi_{j}\in%
\mathbb{R}
$ defined from $\beta_{j}=\sqrt{I_{j}}\exp(\mathrm{i}\varphi_{j})$.
Introducing these expressions into the mean-field equations (\ref{MFeqs}) in
the stationary limit ($\dot{\beta}_{j}=0$), we get
%
\begin{eqnarray}
\sigma =\bar{\beta}_{\mathrm{p}}+\frac{1}{2}\bar{\beta}_{\mathrm{s}}%
^{2}+\frac{g^{2}\bar{\beta}_{\mathrm{p}}\left[  \bar{I}_{\mathrm{p}}%
-(1+\kappa)(1+\bar{I}_{\mathrm{s}})(1+\kappa+\bar{I}_{\mathrm{s}})\right]
}{4\left[  \bar{I}_{\mathrm{p}}-(1+\kappa)^{2}\right]  \left[  \bar
{I}_{\mathrm{p}}-(1+\bar{I}_{\mathrm{s}})^{2}\right]  },\label{MFpS}
\\
\bar{\beta}_{\mathrm{p}}\bar{\beta}_{\mathrm{s}}^{\ast}  =\left(
1+\frac{g^{2}\kappa\bar{I}_{\mathrm{p}}(2+\kappa+\bar{I}_{\mathrm{s}}%
)}{2\left[  \bar{I}_{\mathrm{p}}-(1+\kappa)^{2}\right]  \left[  \bar
{I}_{\mathrm{p}}-(1+\bar{I}_{\mathrm{s}})^{2}\right]  }\right)  \bar{\beta
}_{\mathrm{s}}, \label{MFsS}
\end{eqnarray}
%
where the bar denotes `steady-state values'.%

It is then straightforward to show that the pump phase $\bar{\varphi
}_{\mathrm{p}}$ is locked to $0$, while the signal phase $\bar{\varphi
}_{\mathrm{s}}$ can take the values $0$ or $\pi$, just as in the classical
solution. As for the intensities, equation (\ref{MFsS}) yields a third order
polynomial in $I_{\mathrm{s}}$, with a trivial root $\bar{I}_{\mathrm{s}}=0$
(below threshold solution) and second root (above threshold solution) which
can be written in terms of the pump intensity as%
\begin{equation}
\bar{I}_{\mathrm{s}}=\frac{\sqrt{\bar{I}_{\mathrm{p}}R(\bar{I}_{\mathrm{p}}%
)}+\kappa g^{2}\bar{I}_{\mathrm{p}}}{4\left(  \sqrt{\bar{I}_{\mathrm{p}}%
}-1\right)  \left[  (1+\kappa)^{2}-\bar{I}_{\mathrm{p}}\right]  }-1, \label{IsAbove}%
\end{equation}
with%
\begin{equation}
\fl R(\bar{I}_{\mathrm{p}})=\left\{  4\left(  \sqrt{\bar{I}_{\mathrm{p}}%
}-1\right)  \left[  (1+\kappa)^{2}-\bar{I}_{\mathrm{p}}\right]  -\kappa
(1+\kappa)g^{4}\right\}  ^{2}-\kappa^{2}g^{4}\left[  (1+\kappa)^{2}-\bar
{I}_{\mathrm{p}}\right]  ;
\end{equation}
the third root is not relevant, since it can be checked that it leads to
unphysical results, as commented below. Finally, equation (\ref{MFpS}) gives
an equation for the pump intensity, which can be written as%
\begin{equation}
\sigma=\frac{1}{2}\bar{I}_{\mathrm{s}}+\sqrt{\bar{I}_{\mathrm{p}}}\left\{
1+\frac{g^{2}}{4}\frac{\bar{I}_{\mathrm{p}}-(1+\kappa)(1+\bar{I}_{\mathrm{s}%
})(1+\kappa+\bar{I}_{\mathrm{s}})}{\left[  (1+\kappa)^{2}-\bar{I}_{\mathrm{p}%
}\right]  \left[  \bar{I}_{\mathrm{p}}-(1+\bar{I}_{\mathrm{s}})^{2}\right]
}\right\}  ; \label{IpSCEq}%
\end{equation}
below threshold ($\bar{I}_{\mathrm{s}}=0$), this gives a third order
polynomial in $\sqrt{\bar{I}_{\mathrm{p}}}$ whose roots can be found
analytically (although not much insight is gained from their complicated
expression, so we don't give them explicitly), but above threshold it gives a
high-order polynomial whose roots we've only been able to find numerically. Of
the many solutions for $\bar{I}_{\mathrm{p}}$ obtained from this equation,
most of them are disregarded because they are either negative or complex, make
$\bar{I}_{\mathrm{s}}$ in (\ref{IsAbove}) negative or complex, or lead to
negative photon numbers $\langle\hat{a}_{j}^{\dagger}\hat{a}_{j}\rangle$ or
quadrature fluctuations incompatible with the Heisenberg uncertainty principle
$\langle\delta\hat{x}_{j}^{2}\rangle\langle\delta\hat{y}_{j}^{2}\rangle\geq1$
(see below for a definition of the quadrature fluctuations). It is quite
remarkable that, after discarding all these unphysical solutions, only two
solutions for $\bar{I}_{\mathrm{p}}$ remain: one of the three roots of the
below-threshold polynomial---(\ref{IpSCEq}) with $\bar{I}_{\mathrm{s}}=0$---,
and another one from the above threshold one---(\ref{IpSCEq}) with $\bar
{I}_{\mathrm{s}}$ given by (\ref{IsAbove})---.

In Fig. \ref{Fig1ab} we plot the pump (a) and signal (b) intensities
associated to these solutions as a function of the injection; it is apparent
that they tend to the classical solutions far away from the critical point,
but never reach, in particular, the value of the intensities that makes the
linear stability matrix become singular $(\bar{I}_{\mathrm{s}}=0,\bar
{I}_{\mathrm{p}}=1)$. Hence, the solutions of our \textit{self-consistent
linearization} can be seen as regularized versions of the classical below- and
above-threshold solutions, which remove the divergences of the linear theory
of quantum fluctuations (see Section \ref{QuantumResults} for\ and explicit
discussion of the regularized quantum properties). The figures also allow us
to see the way in which this regularization occurs: in the case of the
below-threshold solution (solid blue curves), the pump intensity does not grow
quadratically with the injection as happens with the classical solution
(thin-solid light-grey curve), staying below its critical value $I_{\mathrm{p}%
}=1$ for any physical value of the injection $\sigma$; as for the
above-threshold solution (dashed-dotted red curve), instead of connecting
continuously with the below-threshold solution as the classical solution does,
its appearance is delayed a little bit respect to the classical threshold
$\sigma=1$, and starts with a nonzero signal intensity $I_{\mathrm{s}}$. It
might seem strange that the above- and below-threshold solutions do not
connect continuously, but in Section \ref{QuantumResults} we will argue why
the presence or not of this `jump' is quite irrelevant indeed, as can be
intuitively understood from the point of view of the symmetry breaking that
occurs above threshold: too close to threshold quantum tunneling between the
solutions with opposite signal phase is too fast, and it makes no sense to
talk about these solutions independently.

In summary, in this section we have provided a self-consistent linearized
theory for the DOPO in which the mean-field amplitudes are found not from the
classical nonlinear equations of motion, but from ones including quantum
corrections; this method provides regularized versions of the solutions that
would be obtained from the bare classical theory, which avoid in particular
the critical values of the mean-field amplitudes at which the linear stability
matrix becomes singular, thus avoiding the unphysical results related to it.
Let us remark that this self-consistent method that we have put forward was
already introduced in \cite{Veits97} for the DOPO problem, but only below
threshold; moreover, in the next section we give full meaning to the method by
showing that, within the Schr\"{o}dinger picture, it is equivalent to making a
general Gaussian-state ansatz consistent with the master equation.

\section{Schr\"{o}dinger picture approach: general Gaussian
ansatz\label{Schrodinger}}

The self-consistent linearization introduced in the previous section admits a
simple interpretation from the point of view of the state of the system: we
argue in the following that it is equivalent to making a general Gaussian
ansatz for it. In order to show this, our starting point is the master
equation (\ref{MasterEq}), which in terms of the normalized parameters
(\ref{NormPars}) and variables (\ref{NormVars}) can be rewritten as%
\begin{equation}
g^{2}\frac{d\hat{\rho}}{d\tau}=\left[  \hat{A},\hat{\rho}\right]
+\sum_{j=\mathrm{p},\mathrm{s}}(2\hat{b}_{j}\hat{\rho}\hat{b}_{j}^{\dagger
}-\hat{b}_{j}^{\dagger}\hat{b}_{j}\hat{\rho}-\hat{\rho}\hat{b}_{j}^{\dagger
}\hat{b}_{j}),
\end{equation}
where we have defined an anti-hermitian operator $\hat{A}=\sigma(\hat
{b}_{\mathrm{p}}^{\dagger}-\hat{b}_{\mathrm{p}})+(\hat{b}_{\mathrm{p}}\hat
{b}_{\mathrm{s}}^{\dagger2}-\hat{b}_{\mathrm{p}}^{\dagger}\hat{b}_{\mathrm{s}%
}^{2})/2$.

The evolution equation for the expectation value of any operator $\hat{B}$ is
obtained as
\begin{equation}
g^{2}\frac{d}{d\tau}\langle\hat{B}\rangle=g^{2}\mathrm{tr}\left\{  \hat
{B}\frac{d\hat{\rho}}{d\tau}\right\}  =\langle\lbrack\hat{B},\hat{A}%
]\rangle+\sum_{j=\mathrm{p},\mathrm{s}}\left(  \langle\lbrack\hat{b}%
_{j}^{\dagger},\hat{B}]\hat{b}_{j}\rangle+\langle\hat{b}_{j}^{\dagger}[\hat
{B},\hat{b}_{j}]\rangle\right)  .
\end{equation}
Applied to the annihilation operators $\hat{b}_{j}$, this equation provides
exactly the mean-field equations (\ref{MFeqs}) found in the previous section
for the amplitudes $\beta_{j}=\langle\hat{b}_{j}\rangle$, which depend on the
second moments $\langle\delta\hat{b}_{\mathrm{s}}^{2}\rangle$ and
$\langle\delta\hat{b}_{\mathrm{p}}\delta\hat{b}_{\mathrm{s}}^{\dagger}\rangle
$. On the other hand, the evolution equations of these second moments depend
on third-order moments, and here is where the Gaussian-state approximation
enters into play: we assume that the state of the system is Gaussian
\cite{Weedbrook12} at all times, meaning that higher order moments can be
written as products of first and second moments only. In particular, this has
the consequence that third order moments of quantum fluctuations vanish, that
is, $\langle\delta\hat{b}_{j}^{m}\delta\hat{b}_{k}^{n}\delta\hat{b}_{l}%
^{p}\rangle=0$, where $m$, $n$, and $p$ can be either dagger or nothing. Under
this assumption, the evolution equations of the second order moments%
\begin{eqnarray}
\fl \mathbf{m}=\mathrm{col}(\langle\delta\hat{b}_{\mathrm{p}}^{2}%
\rangle,\langle\delta\hat{b}_{\mathrm{p}}^{2}\rangle^{\ast},\langle\delta
\hat{b}_{\mathrm{p}}^{\dagger}\delta\hat{b}_{\mathrm{p}}\rangle,\langle
\delta\hat{b}_{\mathrm{p}}\delta\hat{b}_{\mathrm{s}}\rangle,\langle\delta
\hat{b}_{\mathrm{p}}\delta\hat{b}_{\mathrm{s}}\rangle^{\ast},\langle\delta
\hat{b}_{\mathrm{p}}\delta\hat{b}_{\mathrm{s}}^{\dagger}\rangle
\\
\qquad\qquad\qquad\qquad\qquad\qquad\quad,\langle
\delta\hat{b}_{\mathrm{p}}\delta\hat{b}_{\mathrm{s}}^{\dagger}\rangle^{\ast
},\langle\delta\hat{b}_{\mathrm{s}}^{2}\rangle,\langle\delta\hat
{b}_{\mathrm{s}}^{2}\rangle^{\ast},\langle\delta\hat{b}_{\mathrm{s}}^{\dagger
}\delta\hat{b}_{\mathrm{s}}\rangle), \nonumber
\end{eqnarray}
form the following closed linear system%
\begin{equation}
\frac{d\mathbf{m}}{d\tau}=\mathcal{M}(\beta_{\mathrm{p}},\beta_{\mathrm{s}%
})\mathbf{m}+\mathbf{n}(\beta_{\mathrm{p}}), \label{GaussianEqs}%
\end{equation}
with the matrix $\mathcal{M}$ given by%
\[
\fl \left(
\begin{array}
[c]{cccccccccc}%
-2\kappa & 0 & 0 & -2\kappa\beta_{\mathrm{s}} & 0 & 0 & 0 & 0 & 0 & 0\\
0 & -2\kappa & 0 & 0 & -2\kappa\beta_{\mathrm{s}}^{\ast} & 0 & 0 & 0 & 0 & 0\\
0 & 0 & -2\kappa & 0 & 0 & -\kappa\beta_{\mathrm{s}}^{\ast} & -\kappa
\beta_{\mathrm{s}} & 0 & 0 & 0\\
\beta_{\mathrm{s}}^{\ast} & 0 & 0 & -(1+\kappa) & 0 & \beta_{\mathrm{p}} & 0 &
-\kappa\beta_{\mathrm{s}} & 0 & 0\\
0 & \beta_{\mathrm{s}} & 0 & 0 & -(1+\kappa) & 0 & \beta_{\mathrm{p}}^{\ast} &
0 & -\kappa\beta_{\mathrm{s}}^{\ast} & 0\\
0 & 0 & \beta_{\mathrm{s}} & \beta_{\mathrm{p}} & 0 & -(1+\kappa) & 0 & 0 &
0 & -\kappa\beta_{\mathrm{s}}\\
0 & 0 & \beta_{\mathrm{s}}^{\ast} & 0 & \beta_{\mathrm{p}} & 0 & -(1+\kappa) &
0 & 0 & -\kappa\beta_{\mathrm{s}}^{\ast}\\
0 & 0 & 0 & 2\beta_{\mathrm{s}}^{\ast} & 0 & 0 & 0 & -2 & 0 & 2\beta
_{\mathrm{p}}\\
0 & 0 & 0 & 0 & 2\beta_{\mathrm{s}}^{\ast} & 0 & 0 & 0 & -2 & 2\beta
_{\mathrm{p}}^{\ast}\\
0 & 0 & 0 & 0 & 0 & \beta_{\mathrm{s}}^{\ast} & \beta_{\mathrm{s}} &
\beta_{\mathrm{p}}^{\ast} & \beta_{\mathrm{p}} & -2
\end{array}
\right)  ,
\]
and
\begin{equation}
\mathbf{n}=g^{2}\mathrm{col}(0,0,0,0,0,0,0,\beta_{\mathrm{p}}%
,\beta_{\mathrm{p}}^{\ast},0);\end{equation}
it is simple to show that the solutions to this linear system coincide with
the moments obtained from the linearized quantum Langevin equations
(\ref{LinLan}), in particular steady-state moments such as (\ref{SS-moments}),
and hence, the self-consistent linearization introduced in the previous
section is strictly equivalent to the assumption that the state of the system
is a general Gaussian state whose moments satisfy the constrains imposed by
the master equation. Note that, since we have three possible solutions for the
mean-field amplitudes $\beta_{j}$, the below-threshold solution with
$\bar{\beta}_{\mathrm{s}}=0$ and two above-threshold solutions $\bar{\beta
}_{\mathrm{s}}=\pm\sqrt{\bar{I}_{\mathrm{s}}}$, we have three Gaussian
ansatzes that we can use, which we will denote by $\hat{\rho}_{\mathrm{G},0}$
and $\hat{\rho}_{\mathrm{G,}\pm}$, respectively. In the next section we
interpret the meaning of these solutions, as well as the jump observed above
threshold for the signal intensity, which was partially discussed in the
previous section.%

\begin{figure}[t]
\flushright \includegraphics[width=\textwidth]{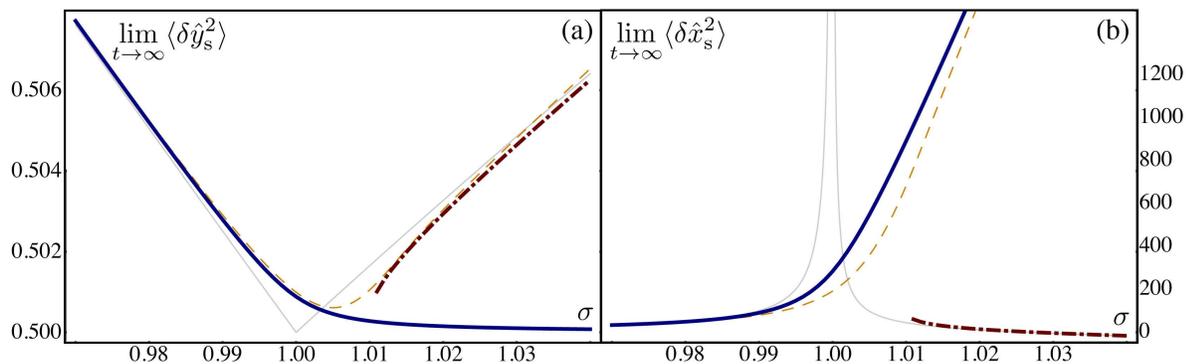}
\caption{Variances of the squeezed (a) and anti-squeezed (b) quadratures as a
function of the injection parameter $\sigma$, for $\kappa=1$ and $g=0.01$
(similar figures are found for any other choice). As in Fig. \ref{Fig1ab}, the
thin-solid light-grey curve corresponds to the usual linearized description;
the solid blue line and dashed-dotted red line correspond to our
self-consistent method below and above threshold, respectively; and the dashed
yellow line corresponds to the predictions of Drummond and collaborators'
perturbative analysis \cite{Kinsler95, Drummond02}.}
\label{Fig1cd}
\end{figure}

\section{Analysis of the results and comparison with previous
methods\label{QuantumResults}}

Let us now analyze the predictions that this self-consistent linear theory
makes for the squeezing of the intracavity signal field, and compare it with
some known beyond-linear results. Let us then define the quadratures $\hat
{x}_{\mathrm{s}}=\hat{a}_{\mathrm{s}}^{\dagger}+\hat{a}_{\mathrm{s}}$ and
$\hat{y}_{\mathrm{s}}=\mathrm{i}(\hat{a}_{\mathrm{s}}^{\dagger}-\hat
{a}_{\mathrm{s}})$, and the corresponding fluctuation operators, $\delta
\hat{x}_{\mathrm{s}}=\hat{x}_{\mathrm{s}}-\langle\hat{x}_{\mathrm{s}}\rangle$
and $\delta\hat{y}_{\mathrm{s}}=\hat{y}_{\mathrm{s}}-\langle\hat
{y}_{\mathrm{s}}\rangle$; it is simple from (\ref{LinLan}) or
(\ref{GaussianEqs}) to obtain the following expression for their variance in
the steady-state:
%
\begin{eqnarray}
\lim_{t\rightarrow\infty}\langle\delta\hat{x}_{\mathrm{s}}^{2}\rangle &
=\frac{(1+\kappa)(1+\bar{I}_{\mathrm{s}})-\sqrt{\bar{I}_{\mathrm{p}}}}{\left(
1+\bar{I}_{\mathrm{s}}-\sqrt{\bar{I}_{\mathrm{p}}}\right)  \left(
1+\kappa-\sqrt{\bar{I}_{\mathrm{p}}}\right)  },
\\
\lim_{t\rightarrow\infty}\langle\delta\hat{y}_{\mathrm{s}}^{2}\rangle &
=\frac{(1+\kappa)(1+\bar{I}_{\mathrm{s}})+\sqrt{\bar{I}_{\mathrm{p}}}}{\left(
1+\bar{I}_{\mathrm{s}}+\sqrt{\bar{I}_{\mathrm{p}}}\right)  \left(
1+\kappa+\sqrt{\bar{I}_{\mathrm{p}}}\right)  }. \nonumber
\end{eqnarray}
%
When the classical solutions for $\bar{I}_{\mathrm{p}}$ and $\bar
{I}_{\mathrm{s}}$ are considered, one immediately sees that $\langle\delta
\hat{x}_{\mathrm{s}}^{2}\rangle\rightarrow\infty$ and $\langle\delta\hat
{y}_{\mathrm{s}}^{2}\rangle\rightarrow0.5$ at threshold $\sigma=1$, which is
exactly the unphysical prediction that we were talking about, since infinite
quadrature fluctuations imply infinite photon number, that is, $\langle\hat
{a}_{\mathrm{s}}^{\dagger}\hat{a}_{\mathrm{s}}\rangle\rightarrow\infty$. On
the other hand, when we introduce in these expressions the regularized
solutions discussed in Section \ref{Heisenberg}, obtained through the
self-consistent method, $\langle\delta\hat{x}_{\mathrm{s}}^{2}\rangle$ becomes
finite in all parameter space, while at the same time, the squeezing level of
$\langle\delta\hat{y}_{\mathrm{s}}^{2}\rangle$ is reduced; this can be
appreciated in Fig. \ref{Fig1cd}. More quantitatively, exactly at the critical
point $\sigma=1$, it is simple to obtain the following expressions for the
quadrature variances to the leading order in $g$:%
\begin{equation}
\lim_{t\rightarrow\infty}\langle\delta\hat{x}_{\mathrm{s}}^{2}\rangle
\approx\frac{2\sqrt{2}}{g},\quad\mathrm{and}\quad\lim_{t\rightarrow\infty
}\langle\delta\hat{y}_{\mathrm{s}}^{2}\rangle\approx0.5+\frac{g}{8\sqrt{2}},
\label{AppVars}%
\end{equation}
showing explicitly how the anti-squeezing is regularized, while the variance
of the squeezed quadrature is increased with respect to its value obtained with the
usual linearization procedure.%

\begin{figure}[t]
\flushright \includegraphics[width=0.9\textwidth]{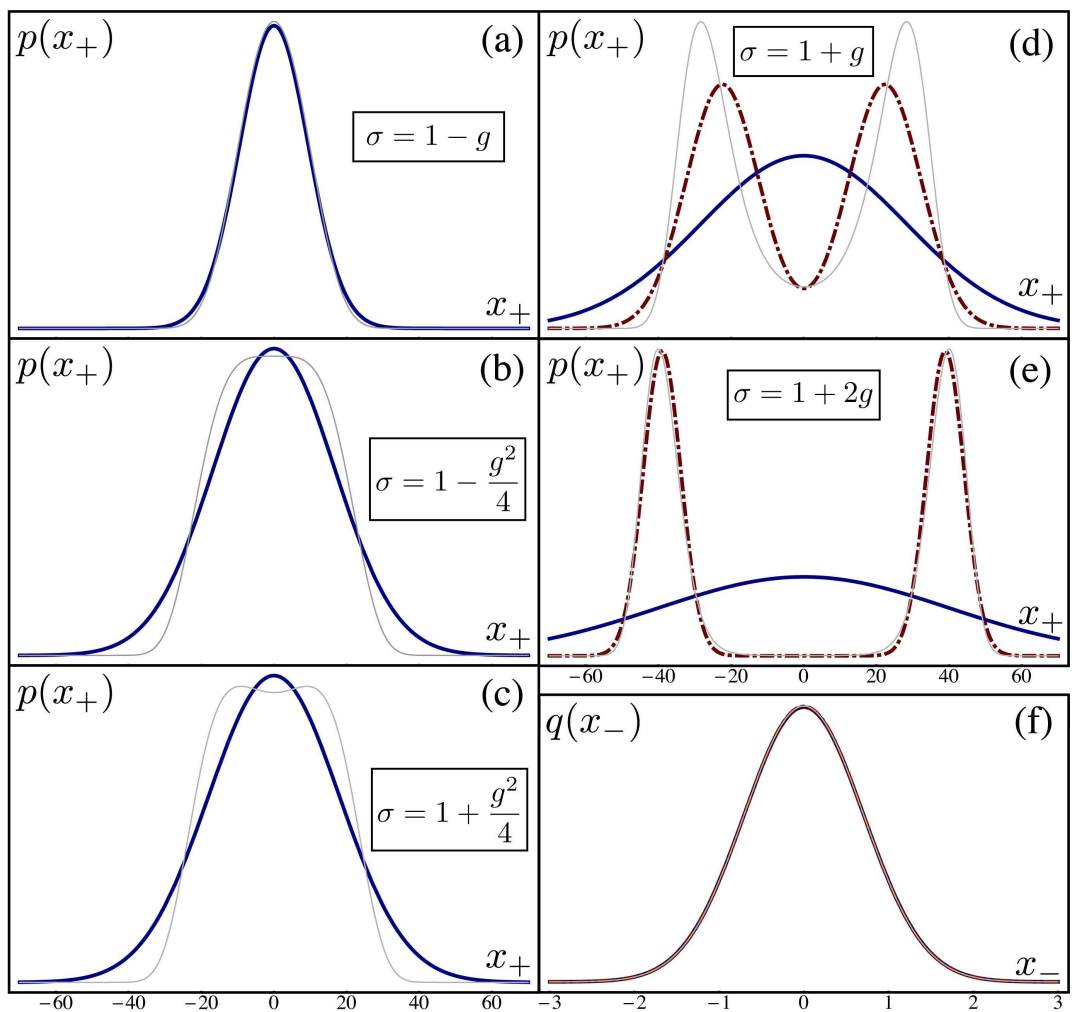}
\caption{Marginal $p(x_{+})$ corresponding to the positive $P$ distribution of
the signal field in the $\kappa\rightarrow\infty$ limit as obtained from the
exact solution (thin-solid light-grey), and our Gaussian ansatzes below (solid blue)
and above (dashed-dotted red) threshold. We have picked the value
$g=0.01$ and show five values of $\sigma$: $1-g$ (a), $1-g^{2}/4$ (b),
$1+g^{2}/4$ (c), $1+g$ (d), and $1+2g$ (e). In (f) we show the marginal
$q(x_{-})$ for $\sigma=1+g$; in the case of this marginal, similar figures are
found for any other value of $\sigma$ around the classical threshold.}
\label{Fig2}
\end{figure}

In order to understand how good the self-consistent linearization is from a
quantitative point of view, we now compare these results with the ones
obtained from the perturbative approach that Drummond and collaborators
developed in the vicinities of the critical point, by making a consistent
multiple-scale expansion of the stochastic variables within the positive $P$
representation \cite{Kinsler95, Drummond02}. This procedure has the virtue of
being valid for any $\kappa$, although it is reliable only close enough to
threshold, concretely for $\left\vert \sigma-1\right\vert <g/\sqrt{2}$;
nevertheless, since we are mainly interested in how the self-consistent
linearization regularizes the conventional linearization around the critical
point $\sigma=1$, where the divergences appear, this will be enough for
comparing with our results. For our purposes, their most relevant results
concern the steady-state (normalized) pump amplitude and the quadrature
fluctuations of the signal field, which read
%
\begin{eqnarray}
\label{Drummond}
\lim_{t\rightarrow\infty}\langle\hat{b}_{\mathrm{p}}\rangle &  \approx
\sigma-\frac{g}{4\sqrt{2}}\langle x^{2}\rangle,\label{Drummond1}
\\
\lim_{t\rightarrow\infty}\langle\delta\hat{x}_{\mathrm{s}}^{2}\rangle &
\approx\frac{\sqrt{2}}{g}\langle x^{2}\rangle, \quad \lim_{t\rightarrow\infty}\langle\delta\hat{y}_{\mathrm{s}}^{2}\rangle &
\approx\frac{3-\sigma}{4}+\frac{g}{16\sqrt{2}}\left(  \frac{2+3\kappa
}{2+\kappa}\right)  \langle x^{2}\rangle, \label{Drummond3}
\end{eqnarray}
%
to the leading order in $g$, where we have defined a real stochastic variable
$x$ distributed according to the probability density function%
\begin{equation}
D(x)=d\exp\left[  \frac{\sigma-1}{\sqrt{2}g}x^{2}-\frac{x^{4}}{16}\right]  ,
\end{equation}
with $d$ a suitable normalization constant. The square of (\ref{Drummond1})
corresponds to the curve that we plotted in Fig. \ref{Fig1ab}(a) to compare
with the steady-state intensity $\bar{I}_{\mathrm{p}}$ that our
self-consistent linearization provides. On the other hand, exactly at
threshold ($\sigma=1$) the moments derived from this distribution admit very
simple expressions in terms of Gamma functions $\Gamma(z)$, and in particular
we have $\langle x^{2}\rangle=4\Gamma(3/4)/\Gamma(1/4)\approx 4/3$, what, together with
(\ref{AppVars}), allows us to write%
%
\begin{eqnarray}
\frac{\left.  \lim_{t\rightarrow\infty}\langle\delta\hat{x}_{\mathrm{s}}%
^{2}\rangle\right\vert _{\mathrm{Drummond}\;\mathrm{et} \: \mathrm{al}}}{\left.  \lim_{t\rightarrow\infty}\langle\delta\hat{x}_{\mathrm{s}}^{2}\rangle\right\vert_\mathrm{self-consistent}}\approx\frac{2}{3},
\\
\frac{\left.  \lim_{t\rightarrow\infty}\langle\delta\hat{y}_{\mathrm{s}}%
^{2}\rangle\right\vert _{\mathrm{Drummond}\;\mathrm{et} \: \mathrm{al}}-0.5}{\left.  \lim_{t\rightarrow\infty}\langle\delta\hat
{y}_{\mathrm{s}}^{2}\rangle\right\vert _{\mathrm{self-consistent}}-0.5}\approx\frac{2}{3}\left(  1+\frac{2\kappa}{2+\kappa}\right)  .
\end{eqnarray}
%
From these expressions we see that the self-consistent linearization provides
a regularization which compares pretty well with Drummond's predictions, at
least regarding the order of magnitude. In particular, we bring the reader's
attention to the anti-squeezing predicted by the self-consistent method, which
is only 50\% above the one predicted by Drummond and collaborators.

As a last test, we now compare the Gaussian ansatzes proportioned by our
self-consistent linearization against the exact state of the signal field,
which is known in the limit $\kappa\gg1$ (in which the pump field can be
adiabatically eliminated) \cite{Drummond80, Wolinsky88, WallsMilburnBook}; in
particular, the positive $P$ distribution associated to the reduced steady
state of the signal mode is given by \cite{Wolinsky88, WallsMilburnBook}%
\begin{equation}
\fl P(\alpha_{\mathrm{s}},\alpha_{\mathrm{s}}^{+})=\left\{
\begin{array}
[c]{cc}%
K\left[  \left(  \alpha_{\mathrm{s}}^{2}-\frac{2\sigma}{g^{2}}\right)
\left(  \alpha_{\mathrm{s}}^{+2}-\frac{2\sigma}{g^{2}}\right)  \right]
^{-1+2/g^{2}}e^{2\alpha_{\mathrm{s}}\alpha_{\mathrm{s}}^{+}} & \mathrm{for}\;\left\vert
\alpha_{\mathrm{s}}\right\vert ,\left\vert \alpha_{\mathrm{s}}^{+}\right\vert
\leq\sqrt{2\sigma}/g\\
0 & \mathrm{for}\;\left\vert \alpha_{\mathrm{s}}\right\vert ,\left\vert
\alpha_{\mathrm{s}}^{+}\right\vert >\sqrt{2\sigma}/g
\end{array}
\right.  ,
\end{equation}
where $\alpha_\mathrm{s}$ and $\alpha^+_\mathrm{s}$ are real, and $K$ is a suitable normalization factor. Explicit expressions of the
state $\hat{\rho}$ can be built from such a positive $P$ distribution, but,
for our purposes, we only need the fact that it allows for the evaluation of
steady-state moments in normal order as%
\begin{equation}
\lim_{t\rightarrow\infty}\langle\hat{a}_{\mathrm{s}}^{\dagger m}\hat{a}_{\mathrm{s}}^{n}\rangle=\int_{%
\mathbb{R}
^{2}}d\alpha_{\mathrm{s}}d\alpha_{\mathrm{s}}^{+}P(\alpha_{\mathrm{s}%
},\alpha_{\mathrm{s}}^{+})\alpha_{\mathrm{s}}^{+m}\alpha_{\mathrm{s}}^{n}.
\end{equation}
Let us define the variables $x_{+}=\alpha_{\mathrm{s}}+\alpha_{\mathrm{s}}%
^{+}$ and $x_{-}=\alpha_{\mathrm{s}}-\alpha_{\mathrm{s}}^{+}$, noting that
$x_{+}$ corresponds directly to the stochastic representation of the anti-squeezed quadrature $\hat
{x}_{\mathrm{s}}=\hat{a}_{\mathrm{s}}+\hat{a}_{\mathrm{s}}^{\dag}$,
while $x_{-}$ corresponds to `$\mathrm{i}$-times' the squeezed
one, that is, to $\mathrm{i}\hat{y}_{\mathrm{s}}=\hat{a}_{\mathrm{s}%
}-\hat{a}_{\mathrm{s}}^{\dag}$. In Figs. \ref{Fig2}(a-e) we show how the
marginal $p(x_{+})=\int_{%
\mathbb{R}
}dx_{-}P(x_{+},x_{-})$ changes as we cross the threshold. The complementary
marginal $q(x_{-})=\int_{%
\mathbb{R}
}dx_{+}P(x_{+},x_{-})$ is shown only at one value of $\sigma$ because it does
not change perceptibly around the critical point, and the Gaussian ansatzes
adapt to it almost perfectly, as can be appreciated in Fig. \ref{Fig2}(f).

The first physically relevant result that this exact steady-state solution predicts is
$\langle\hat{a}_{\mathrm{s}}\rangle=0$ for all $\sigma$. This might seem
surprising, since it seems to suggest that the signal field is never switched
on, that is, that the above-threshold solution with $\langle\hat
{a}_{\mathrm{s}}\rangle\neq0$ characteristic of DOPOs is incorrect; this,
however, is not true, the right answer is a bit more subtle: as $\sigma$ is
increased, the distribution develops two peaks (see Fig. \ref{Fig2}) which,
individually, correspond to the above-threshold amplitudes $\bar{\beta
}_{\mathrm{s}}=\pm\sqrt{\bar{I}_{\mathrm{s}}}$, but, since they are developed
together, their contributions to $\langle\hat{a}_{\mathrm{s}}\rangle$ average
to zero. From a more fundamental point of view, this just reflects the fact
that the master equation (\ref{MasterEq}) has the $%
\mathbb{Z}
_{2}$ symmetry $\hat{a}_{\mathrm{s}}\rightarrow-\hat{a}_{\mathrm{s}}$, and
hence, if $\hat{\rho}_{\mathrm{sol,+}}$ is a symmetry-breaking ansatz with
$\langle\hat{a}_{\mathrm{s}}\rangle=\sqrt{\bar{I}_{\mathrm{s}}/g}$, then
$\hat{\rho}_{\mathrm{sol,-}}=\exp(-\mathrm{i}\pi\hat{a}_{\mathrm{s}}^{\dagger
}\hat{a}_{\mathrm{s}})\hat{\rho}_{\mathrm{sol,+}}\exp(\mathrm{i}\pi\hat
{a}_{\mathrm{s}}^{\dagger}\hat{a}_{\mathrm{s}})$, which has $\langle\hat
{a}_{\mathrm{s}}\rangle=-\sqrt{\bar{I}_{\mathrm{s}}/g}$, is another equally
valid ansatz; in the absence of any bias, the state of the system has to be
regarded as the balanced mixture $\hat{\rho}_{\mathrm{sol}}=(\hat{\rho
}_{\mathrm{sol,+}}+\hat{\rho}_{\mathrm{sol,-}})/2$, which is the one giving
the correct experimental statistics: every time the DOPO is switched on, it
has to choose the phase $0$ or $\pi$ according to the particular initial
fluctuations from which the steady state is built up, that is, according to
\textit{spontaneous symmetry breaking}. It feels natural to think that one can
force the system to pick one particular phase at every experimental run (at
least in a metastable sense) by adding an explicit symmetry breaking mechanism
such as the injection of a weak laser at the signal frequency---a term like
$\hat{H}_{\mathrm{inj,s}}=\mathrm{i}\hbar\mathcal{E}_{\mathrm{s}}(\hat
{a}_{\mathrm{s}}^{\dagger}-\hat{a}_{\mathrm{s}})$ in the Hamiltonian---;
however, this picture is only correct once enough above threshold, since close
to threshold quantum tunneling between the states $\hat{\rho}_{\mathrm{sol,\pm
}}$ is too fast \cite{Drummond80,Drummond89,Kinsler91}, and no phase locking can be achieved within the observation
time. In other words, from an observational point of view, it only makes sense
to analyze the properties of $\hat{\rho}_{\mathrm{sol,+}}$ and $\hat{\rho
}_{\mathrm{sol,-}}$ separately once the peaks of the positive $P$ distribution
are enough far apart, and this is why we made the statement that the jump seen
in the signal intensity above threshold with the self-consistent method is not
relevant for real applications, as it just reflects the fact that some
distance from threshold is required for $\hat{\rho}_{\mathrm{sol,\pm}}$ to
have independent meaning, as otherwise fast tunneling times
prevent their existence.

Knowing the exact form of the positive $P$ distribution in the limit
$\kappa\gg1$ allows us to get a pictorial feeling of how good our Gaussian
ansatzes are. In particular, superposed to the exact marginals (solid-thin
light-grey line), in Fig. \ref{Fig2} we plot the marginals corresponding to
the Gaussian ansatzes $\hat{\rho}_{\mathrm{G},0}$ (solid blue) and $\hat
{\rho}_{\mathrm{G},>}=(\hat{\rho}_{\mathrm{G},+}+\hat{\rho}_{\mathrm{G},-})/2$
(dashed-dotted red). We can appreciate how $\hat{\rho}_{\mathrm{G},0}$ adapts
very well to $\hat{\rho}_{\mathrm{sol}}$ below threshold, except close to the
point in which the peaks start developing ($\sigma=1-g^{2}/4$), where the
exact distribution flattens in the center, loosing its approximate
Gaussianity. Note also that the above-threshold Gaussian ansatz $\hat{\rho
}_{\mathrm{G},>}$ appears at $\sigma=1+g$ in this $\kappa\rightarrow\infty$
limit, and it quite rapidly converges to the exact $\hat{\rho}_{\mathrm{sol}}$
as we increase $\sigma$.

\section{Conclusions and outlook}

In summary, we have developed a linearization procedure for nonlinear optical
cavities in which the mean-field amplitudes are found self-consistently by
introducing some information about the quantum fluctuations in their evolution
equations; we have applied it to one of the simplest nonlinear resonators, the
DOPO, showing how the method is capable of regularizing the divergences found
at the critical point with the traditional linearized analysis. We have also
shown that such procedure is completely equivalent to using a general Gaussian
ansatz for the state of the system. Finally, we have compared the results
derived from the self-consistent linearization with other known exact
(positive $P$ distribution under adiabatic elimination of the pump
\cite{Wolinsky88, WallsMilburnBook}) or quasi-exact (Drummond and
collaborators' perturbative expansion \cite{Kinsler95, Drummond02}) methods
for the DOPO, proving that they are reliable qualitatively, and also roughly
quantitatively, with the advantage that two-time correlators (and hence output
spectra of observables) are straightforwardly evaluated within the linearized
description, what we think will be useful to study quantum correlations in
more complicated systems such as non-degenerate or multi-mode OPOs
\cite{Reid88, Drummond90,Navarrete08,Navarrete09}. The application of the
method to systems with other types of critical points and bifurcations is an
open question that will be interesting to analyze in the future as well; the
bistability found in Kerr resonators \cite{DrummondKerr80,Vogel88} or the Hopf
bifurcations found in optomechanical cavities \cite{Fabre94} and intracavity
second-harmonic generation \cite{Drummond80bis, Lugiato83}, are good candidate scenarios.

\ack
The authors thank Peter D. Drummond, Germ\'{a}n J. de Valc\'{a}rcel, and
G\'{e}za Giedke for fruitful discussions. CN-B acknowledges funding from the
Alexander von Humbolt Foundation through their Fellowship for Postdoctoral
Researchers. ER acknowledges support from the Spanish Gorverment through
project FIS2011-26960 and grant PRX12/00264 of the \textquotedblleft Programa
de movilidad de profesores e investigadores seniores en centros extranjeros de
ense\~{n}anza superior e investigaci\'{o}n\textquotedblright. YC and TS are
supported by the EU under the IP project AQUTE.
\\

\end{document}